\documentclass[prb,twocolumn,showpacs,preprintnumbers,amsmath,amssymb]{revtex4}

\usepackage{epsfig}
\usepackage{graphicx}
\usepackage{dcolumn}
\usepackage{bm}

\renewcommand{\l}{\lambda}
\newcommand{\I}{\text{i}}
\renewcommand{\k}{{\bf k}}
\newcommand{\q}{{\bf q}}
\renewcommand{\d}{{\bm \delta}}
\newcommand{\tmu}{\tilde\mu}

\def\lsim{\lower.35em\hbox{$\stackrel{\textstyle<}{\textstyle\sim}$}}

\begin{document}
\title{Analytical expressions for the polarizability of the honeycomb lattice}
\author{T. Stauber}

\affiliation{Dep. de F\'{\i}sica de la Materia Condensada, Universidad Aut\'onoma de Madrid, E-28049 Madrid, Spain}
\date{\today}

\begin{abstract}
We present analytical expressions for the polarizability $P_\mu(q_x,\omega)$ of graphene modeled by the hexagonal tight-binding model for small wave number $q_x$, but arbitrary chemical potential $\mu$. Generally, we find $P_\mu(q_x,\omega)=P_\mu^<(\omega/\omega_q^{})+q_x^2P_\mu^>(\omega)$ with $\omega_q^{}=v_F^{}q_x$ the Dirac energy, where the first term is due to intra-band and the second due to inter-band transitions. Explicitly, we derive the analytical expression for the imaginary part of the polarizability including intra-band contributions and recover the result obtained from the Dirac cone approximation for $\mu\rightarrow0$. For $\mu<\sqrt{3}t$, there is a square-root singularity at $\omega_q^{}=v_F^{}q_x$ independent of $\mu$. For doping levels close to the van Hove singularity, $\mu=t\pm\delta\mu$, ${\rm Im}P_\mu(q_x,\omega)$ is constant for $\delta\mu/t<\omega/\omega_q^{}\ll1$.
\end{abstract}

\maketitle
\section{Introduction}
Graphene is a two-dimensional carbon allotrope which has attracted immense research activity due to its novel mechanical and electronic properties.\cite{Geim09,Neto09,DasSarma10,Peres10} It is also interesting in view of potential applications in nanomechanical and nanoelectronic devices and one of the reasons for this lies in the possibility to change the carrier density and type by applying a gate voltage between graphene and the isolating substrate. This gave rise to the celebrated ambipolar field effect with an almost constant mobility of around $10^4$ cm${}^2$/Vs for graphene on SiO${}_2$-substrate.\cite{Nov04} 

But it is also possible to modify the Fermi surface via chemical doping, i.e., by the deposition of e.g. potassium atoms which donate their lone valence electrons to the surface layer.\cite{Ohta} Whereas with a typical back-gate voltage, chemical potentials of the order of a few 100 meV can be reached, doping levels up to the van Hove singularity were recently achieved via chemically $n$-type doping using various combinations of K and Ca on both sides of graphene.\cite{Rotenberg10}

For small energies and chemical potentials $\omega,\mu\lsim1{\rm eV}$, graphene can be well described by the Dirac cone approximation, optionally including weak trigonal warping corrections.\cite{Bostwick07,Falko08} For larger energies or doping levels around the van Hove singularity, the linearization around the $K$-pionts is not applicable anymore and one usually has to resort to numerical methods.\cite{Louie09} Only around special symmetry points like the $M$-point, analytical solutions are still feasible.\cite{Gonzalez08,Stauber10}

The tight-binding model on a hexagonal lattice with only one orbital is a good approximation to the electronic structure of graphene for doping levels up to the $M$-point.\cite{Satpathy09} Including interaction effects via exchange self-energy corrections preserve the trigonal warping of the Fermi surface topology\cite{Roldan08} and only close to the van Hove singularity the local Coulomb interaction $U$ can lead to instabilities at low energies.\cite{Gonzalez03} Superconductivity mediated via electron-electron interaction was hence predicted for graphene provided that the Fermi surface is close to the $M$-point.\cite{Rotenberg10} 

Instabilities due to Coulomb interaction can be analyzed in terms of the effective interaction vertex which is dressed by the static polarizability as a first approximation.\cite{Scalapino87} Usually this is done numerically and only for gated graphene with small chemical potential, analytical expressions were obtained within the Dirac cone approximation.\cite{Shung86,Gon99,Ando06,Wunsch06,Hwang07} Here, we will report on an analytical solution for the imaginary part of the polarizability for arbitrary chemical potential within the tight-binding model. The solution is valid for low energies $\hbar\omega\lsim2\mu$ and small incoming wave vector $|\q|=q_x$ where the $x$-axis denotes the high-symmetry direction which connects the $\Gamma$- and the $M$-point. Analytical expressions are especially important close to the van Hove singularity since numerical methods tend to fail in this regime. They also provide insight in the scaling properties and can explain the peak-splitting in arbitrary direction of the square-root singularity seen in the numerical solution of Ref. \cite{Stauber10}.

\section{Polarizability of the honeycomb lattice}
The density-density correlation or Lindhard function of the honeycomb lattice is given by 
\begin{align}\notag
P_\mu(\q,\omega)&=\frac{-g_s}{(2\pi)^2}\int_{\text{1.BZ}}d^2k\sum_{\l,\l'=\pm}F_{\l,\l'}(\k,\q)\\\label{densitycorrelation}
F_{\l,\l'}(\k,\q)&=f_{\l\cdot \l'}(\k,\q)
\frac{n_F(E^{\l}({\bf k}))-n_F(E^{\l'}({\bf k}+{\bf q}))}{E^{\l}({\bf k})-E^{\l'}({\bf k}+{\bf q})+\hbar\omega+\I0}\;,
\end{align}
with the eigenenergies $E^{\pm}({\bf k})=\pm t|\phi_\k|$ ($t\approx2.7$eV is the hopping amplitude), $n_F(E)$ the Fermi function, $g_s=2$ the spin-degeneracy and $\phi_\k=\sum_\d e^{\I\k\cdot\d}$ the complex structure factor where $\d$ denote the three nearest neighbor vectors. Due to the two gapless bands, the above expression contains the band-overlap function 
\begin{align}
\label{overlapdensity}
f_{\pm}(\k,\q)&=\frac{1}{2}\left(1\pm\text{Re}\left[\frac{\phi_\k}{|\phi_\k|}\frac{\phi_{\k+\q}^*}{|\phi_{\k+\q}|}\right]\right)\;,
\end{align}
which marks the crucial difference to standard text-book results containing only one band.

\begin{figure}[t]
\begin{center}
\includegraphics[angle=0,width=0.8\linewidth]{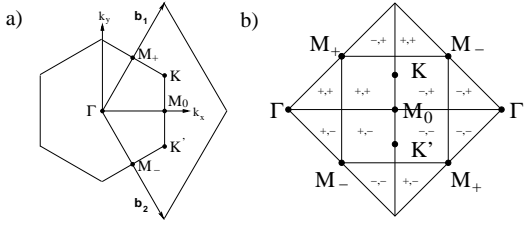}
\caption{a) The hexagonal and rhombical Brillouin zone b) The symmetrized rhombical Brillouin zone and its segmentation. The inner square refers to $j=-$, the outer triangles refer to $j=+$; additionally the values of $s$ and $s'$ are given.} 
  \label{fig:bz}
\end{center}
\end{figure}
It is possible to transform the integration over the rhombical Brillouin zone into an integration over a quadratic area. First, we symmetrize the Brillouin zone by introducing dimensionless parameters $3ak_x/2\rightarrow k_x$ and $\sqrt{3}ak_y/2\rightarrow k_y$, with $a=1.4$\AA$ $ the carbon-carbon distance. Second, the integration over the outer triangles of Fig. \ref{fig:bz} is written as the integration over the quadratic area by symmetrizing the sign of the external wave number $\q\to(sq_x,s'q_y)$ with $s,s'=\pm$. The integration over the inner squares of Fig. \ref{fig:bz} can be written as the integration over the quadratic area by slightly modifying the integrand $F_{\l,\l'}(\k,\q)$. For this we introduce the generalized energy $E^{\l,j}(\k)=\l|\phi_\k^j|$ and overlap function $f_\pm^j$ with
\begin{align}
|\phi_\k^j|=\sqrt{3+2\cos(2k_y)+j4\cos(k_y)\cos(k_x)}\;
\end{align} 
and 
\begin{align}
f_{\pm}^j({\bf k},{\bf q})=\frac{1}{2}\left(1\pm\frac{\tilde f^{j}({\bf k},{\bf q})}{|\phi_\k||\phi_{\k+\q}|}\right)
\end{align}
where
\begin{align}
&\tilde f^{j}({\bf k},{\bf q})=\cos(q_x)+j2\cos(k_y)\cos(k_x+q_x)\\
&+2\left[2\cos(k_y)\cos(q_x/2)+j\cos(k_x+q_x/2)\right]\cos(k_y+q_y)\nonumber
\end{align}
and the dimensionless variables $q_xa\rightarrow q_x$, $\sqrt{3}q_ya/2\rightarrow q_y$.

Changing now the integrand $F_{\l,\l'}(\k,\q)$ to $F_{\l,\l'}^j(\k,\q)$ by substituting $E^\lambda(\k)\to E^{\lambda,j}(\k)$ and $f_\pm(\k,\q)\to f_\pm^j(\k,\q)$, we have
\begin{align}
P_\mu(\q,\omega)&=\frac{g_s\sqrt{3}}{(2\pi)^2}\left(\frac{2}{3a}\right)^2\int_0^{\pi/2} dk_x\int_{0}^{\pi/2} dk_y\notag\\
&\times\sum_{\l,\l'=\pm}\sum_{j=\pm}\sum_{s,s'=\pm}F_{\l,\l'}^j(\k,sq_x,s'q_y)\;.
\end{align}
From this definition of the polarizability, we can make a suitable substitution to obtain an analytical solution.

\section{Analytical solution}
For finite chemical potential $\mu>0$, zero temperature $T=0$ and small wave vector $|\q|a\ll1$, the imaginary part of the polarizability for low energies $\hbar\omega\lsim2\mu$ is only determined by intra-band transitions. This limit shall be denoted by ${\rm Im}P_\mu\rightarrow{\rm Im}P_\mu^<$. With ${\bf q}_{s,s'}=(sq_x,s'q_y)$, $\tilde\mu=\mu/t$, and $\tilde\omega=\hbar\omega/t$, we obtain
\begin{align}
&{\rm Im}P_\mu^<({\bf q},\omega)=\frac{g_s\sqrt{3}}{4\pi t}\left(\frac{2}{3a}\right)^2\int_0^{\pi/2}dk_x\int_0^{\pi/2}dk_y\notag
\sum_{s,s'=\pm}\sum_{j=\pm}\\
\label{ImPM}
&f_+^j({\bf k},{\bf q}_{s,s'}){\bf q}_{s,s'}\nabla_{\bf k}|\phi_{\bf k}^j|
\delta\left(\tilde\mu-|\phi_{\bf k}^j|\right)\delta\left(\tilde\omega-{\bf q}_{s,s'}\nabla_{\bf k}|\phi_{\bf k}^j|\right)\;.
\end{align} 
For $\q$-vectors in $x$-direction, i.e., $q_y=0$, the integral over the two delta-functions can be performed analytically with the substitution $x=\sin k_x$, $y=\sin k_y$. 

The analytical expression depends on the zeros of the first delta function
\begin{align}
y_\pm^2=\frac{3-\tilde\mu^2}{4}\pm\frac{\tilde\mu}{2}\sqrt{1-\left(\omega/\omega_q^{}\right)^2}\;,
\end{align}
where $\omega_q^{}=v_Fq_x$ is the Dirac energy with $v_F=\frac{3}{2}at/\hbar$ the Fermi velocity. The final result can then be written in the compact form
\begin{align}
\label{ImPintra}
{\rm Im}P_\mu^<(q_x,\omega)&=\frac{g_s\sqrt{3}}{8\pi}\frac{\mu}{(\hbar v_F)^2}\frac{\omega}{\sqrt{\omega_q^2-\omega^2}}\\\notag
\times&\theta(\omega_\mu^+-\omega)\left[\frac{\theta(\omega-\omega_\mu^-)}{y_-\sqrt{1-y_-^2}}+\frac{1}{y_+\sqrt{1-y_+^2}}\right]\;,
\end{align}
with $\omega_\mu^-=\theta(\tilde\mu-1)\omega^*$ and $\omega_\mu^+=\omega_q^{}\theta(\sqrt{3}-\tilde\mu)+\omega^*\theta(\tilde\mu-\sqrt{3})$ where $\omega^*=\frac{\omega_q^{}}{2}\sqrt{10-\tilde\mu^2-9/\tilde\mu^2}$. Note that ${\rm Im}P_\mu^<(q_x,\omega)$ only depends on the ratio $\omega/\omega_q^{}$.

Real and imaginary part of a response function are related via the Kramers-Kronig relation. For this, the $\omega$-dependence for the whole spectrum is needed, but we can define the contribution to the real part of the polarizability that originates from intraband transitions: 
\begin{align}
{\rm Re}P_\mu^<(q_x,\omega)=\frac{2}{\pi}\int_0^{\omega_q^{}} d\omega'\frac{\omega'{\rm Im}P_\mu^<(q_x,\omega')}{(\omega')^2-\omega^2}
\end{align}
Also ${\rm Re}P_\mu^<(q_x,\omega)$ depends only on $\omega/\omega_q^{}$. For $\omega=0$, there is no dependence on $q_x$ and we have 
\begin{align}
&{\rm Re}P_\mu^<(q_x,\omega=0)=\rho(\mu)
\end{align}
with $\rho(\mu)=\frac{1}{A}\sum_\k\delta(\mu-t|\phi_\k|)$ the density of states of the tight-binding model which can be expressed in terms of the complete elliptic integral of the first kind.\cite{Castro08}

Let us now discuss the f-sum rule for small $q_x$ defined as\cite{Stauber10b}
\begin{align}
\frac{2}{\pi}\frac{\hbar^2A_c}{(aq_x)^2t}\int_0^{6t} d\omega\omega{\rm Im}P_\mu(q_x,\omega)=\mathcal{I}_\mu\;,
\end{align}
with $\mathcal{I}_\mu=\frac{g_s}{2N}\sum_\k |\phi_\k|\theta(\tilde\mu-|\phi_\k|)$
where $A=A_cN$ is the area of the sample ($A_c=3\sqrt{3}a^2/2$).

Let us divide the total weight $\mathcal{I}_\mu$ into partial weights due to intra-band and inter-band contributions $\mathcal{I}_\mu=\mathcal{I}_\mu^<+\mathcal{I}_\mu^>$. With Eq. (\ref{ImPintra}) and $\tilde\mu\leq1$, $\mathcal{I}_\mu^<$ is then given by
\begin{align}
&\mathcal{I}_\mu^<\!=\frac{9g_s}{4\pi^2}\int_0^1\!\!d\xi\!\!\left[\frac{\sqrt{1-\xi^2}}{\sqrt{(\xi_+-\xi)(\xi_-+\xi)}}+\frac{\sqrt{1-\xi^2}}{\sqrt{(\xi_++\xi)(\xi_--\xi)}}\right],
\end{align}
where $\xi_+=(3-\tilde\mu^2)/2\tmu$ and $\xi_-=(1+\tilde\mu^2)/2\tmu$ which can be expressed in terms of elliptic functions. For $\tilde\mu>1$, the integration bounds have to be modified.

Since $\mathcal I_\mu$ and $\mathcal{I}_\mu^<$ are independent of $q_x$ and the leading order of the band-overlap of the inter-band transition is proportional to $q_x^2$, we have for the inter-band contributions to the polarizibility ${\rm Im}P_\mu^>(q_x,\omega)=q_x^2\theta(\hbar\omega-2\mu)\widetilde P(\omega)$, where $\widetilde P$ only depends on the energy. Obtaining the real part via the Kramers-Kronig relation, we can thus generally write the polarizability as
\begin{align}
P_\mu(q_x,\omega)=P_\mu^<(\omega/\omega_q^{})+q_x^2P_\mu^>(\omega)\;,
\end{align}
where the first term is due to intra-band and the second term due to inter-band processes.

In table \ref{table}, the relative weight of the intra-band transition is shown as they contribute to the total weight.
\begin{table}[h]
\begin{center}
\begin{align}
\notag
\text{
\begin{tabular}{c||c|c|c|c|c|c|c|c}
$\tilde\mu$&0.1&0.5&0.9&1.0&1.1&1.5&2.0&2.5\\
\hline
$\mathcal{I}_\mu$&1.574&1.558&1.462&1.395&1.319&1.076&0.755&0.396\\
\hline
$\mathcal{I}_\mu^</\mathcal{I}_\mu$&0.053&0.266&0.528&0.654&0.773&0.927&0.981&0.997
\end{tabular}}
\end{align}
\caption{Total and relative weight due to intra-band transitions of the f-sum rule for various chemical potentials $\mu$.} 
  \label{table}
\end{center}
\end{table}

\section{Discussion}
\begin{figure}[t]
\begin{center}

\includegraphics[angle=0,width=0.8\linewidth]{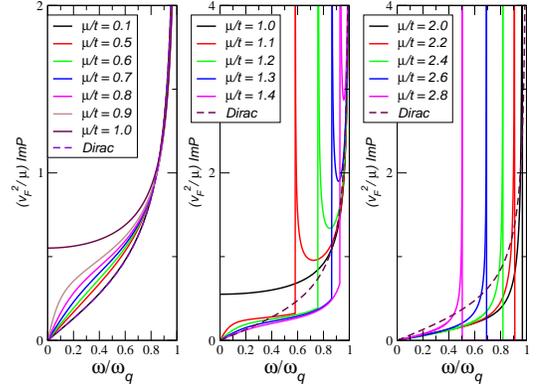}
\caption{(color online): The imaginary part of the polarizability as function of the energy $\omega$ for various chemical potential $\tilde\mu=\mu/t$ divided into the three regimes: $\tilde\mu\leq 1$ (left panel), $1\leq\tilde\mu<\sqrt{3}$ (middle panel), and $\tilde\mu>\sqrt{3}$ (right panel). The result of the Dirac cone approximation is also shown (dashed line).} 
  \label{fig:Pol}
\end{center}
\end{figure}
\begin{figure}[t]
\begin{center}
  \includegraphics[angle=0,width=0.8\linewidth]{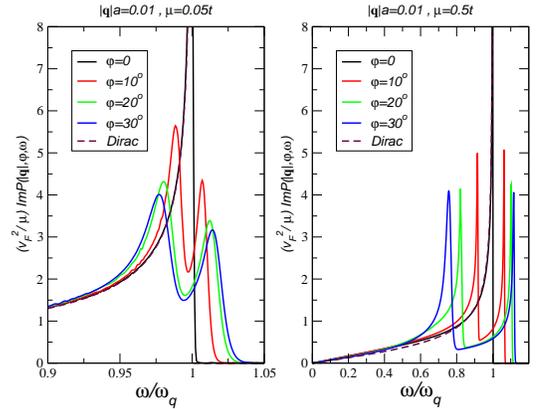}
\caption{(color online): The imaginary part of the polarizability $\text{Im} P^<(|\q|,\varphi,\omega)$ with $|\q|a=0.01$ as function of the energy $\omega$ for various angles $\varphi$ and two chemical potentials $\mu/t=0.05$ (left) and $\mu/t=0.5$ (right) at $k_BT/t=0.01$. Also shown the result from the Dirac cone approximation at zero temperature (dashed line).} 
  \label{fig:Intra1}
\end{center}
\end{figure}
A distinct signature of non-interacting 2D
electrons is a divergent behavior of the polarizability at the threshold
for the excitation of electron-hole pairs determined by the Fermi velocity. For small chemical potential, the Fermi velocity $v_F$ in graphene is independent of $\mu$ and the divergence takes place at $\omega_q^{}=v_Fq_x$. But even in the regime where the Dirac cone approximation does not hold, i.e., curvature in form of trigonal warping has to be taken into account, we find the same divergent behavior at $\omega_q^{}$. It is remarkable that this threshold is independent of the chemical potential $\mu$ up to $\mu<\sqrt{3}t$.  

For $\mu\rightarrow0$, we obtain
\begin{align}
{\rm Im}P_{\mu\to0}^<(q_x,\omega)=\frac{g_s}{\pi}\frac{\mu}{(\hbar v_F)^2}\frac{\omega\theta(\omega_q^{}-\omega)}{\sqrt{\omega_q^2-\omega^2}}\;.
\end{align}
This expression is also obtained from the Dirac cone approximation in the limit $\omega,q\ll\mu$. This shows that the Dirac cone approximation is valid only in the limit $\mu\rightarrow0$ which leads to the criticality of Dirac Fermions at zero gate voltage and zero temperature.\cite{Sachdev08}

Let us now discuss the special case where the chemical potential lies at the van Hove singularity. For $\mu=t$, we have
\begin{align}
\label{PvanHove}
{\rm Im}P_{\mu=t}(q_x,\omega\to0)&=\frac{g_s\sqrt{3}}{2\pi}\frac{t}{(\hbar v_F)^2}\;.
\end{align}

Whereas ${\rm Im}P_{\mu=t}(q_x,\omega\to0)\propto{\rm const}$ is a necessary condition to yield the logarithmic singularity of the density of states at the van Hove singularity, we shall now investigate the behavior for chemical potentials close to the van Hove singularity $\mu=t\pm\delta\mu$. Neglecting quadratic contributions $\mathcal{O}(\delta\mu^2)$ which is valid for $\delta\tilde\mu\ll(\omega/\omega_q^{})$, only the solution $y_+$ contributes to a finite imaginary part for low energies and we obtain
\begin{align}
{\rm Im}P_{\mu=t\pm\delta\mu}(q_x,\omega\to0)&=\frac{g_s\sqrt{3}}{4\pi}\frac{t\pm(\delta\mu/2)}{(\hbar v_F)^2}\;.
\end{align}
We thus find the onset of non-Fermi liquid behavior for excitations in the direction of the highest symmetry for chemical potentials close to the van Hove singularity. For energies with $2\sqrt{\delta\tilde\mu}\ll(\omega/\omega_q^{})\ll1$ and $\mu<t$, also $y_-$ contributes to the constant behavior and we obtain the limit value of Eq. (\ref{PvanHove}). 

In Fig. \ref{fig:Pol}, we show the results in three panels for the three regimes $\tilde\mu\leq 1$ (left panel), $1\leq\tilde\mu<\sqrt{3}$ (middle panel), and $\tilde\mu>\sqrt{3}$ (right panel). For $\tilde\mu<1$, the characteristic feature is the square-root singularity at $\omega=\omega_q^{}$, independent of the $\tilde\mu$. For low chemical potential $\tilde\mu\lsim0.3$, the result agrees well with the Dirac cone approximation (dashed curve). For larger chemical potential, there is increasing weight for lower energies and for $\tilde\mu=1$, $\text{Im}P_\mu$ shows constant behavior for low energies. For chemical potentials with $1<\tilde\mu<\sqrt{3}$, there is - in addition to the square-root singularity at $\omega=\omega_q^{}$ - also a square-root singularity at $\omega=\omega^*$. For $\tilde\mu>\sqrt{3}$, only the square-root singularity at $\omega=\omega^*$ survives. 

For arbitrary direction $\varphi=\tan^{-1}(q_y/q_x)$, in particular for $\q$ in $y$-direction, the solution of the zeros of the delta-functions of Eq. (\ref{ImPM}) involves polynomials of 4th and 6th order and we were not able to obtain an analytical solution. The numerical solution, though, shows the expected peak splitting of the square-root singularity at $\omega=\omega_q^{}$, as shown in Fig. \ref{fig:Intra1}, with the maximal peak-splitting for $\varphi=\pi/6+n\pi/3$, $n\in\mathbb{N}$. As suggested by the analytical expression, the double-peak structure appears even for small chemical potentials, e.g., $\mu/t=0.05$,  for which the Dirac cone approximation should hold. 

The curves of Fig. \ref{fig:Intra1} were obtained for $k_BT/t=0.01$, thus slightly larger than room temperature. The curves for $\varphi=0$ are basically unaffected by temperature, but for arbitrary direction the algebraic divergences seen for $\mu/t=0.5$ are smeared out at larger temperature as it is the case for $\mu/t=0.05$. The curves for $\mu/t=0.05$ and $\varphi\neq0$ develop the algebraic divergence for decreasing temperature, so generally, we can say that the algebraic divergences become broadened when the energy set by the temperature is much larger that the maximal peak-splitting at $\varphi=\pi/6$. For the curves of Fig. \ref{fig:Pol}, we find from the numerical solution that the singularity at $\omega=\omega_q$ is practically unaffected by temperature whereas the singularity at $\omega=\omega^*$ is strongly broadened for $T>0$.

\section{Conclusions}
In summary, we have presented analytical expressions for the imaginary part of the polarizability for wave vectors in the $\Gamma-M$ direction. The results are valid for all chemical potentials $\mu$ for low energies $\hbar\omega\lsim2\mu$ and small wave vector $|\q|a\ll1$. As a special feature, we find a singularity at the Dirac energy $\omega_q^{}=v_Fq$ up to large $\tilde\mu<\sqrt{3}$. For the chemical potential close to the van Hove singularity, we find the onset of non-Fermi liquid behavior with ${\rm Im}P_{\mu=t\pm\delta\mu}\propto{\rm const}$ for $\delta\mu/t<\omega/\omega_q^{}\ll1$. For arbitrary direction $\varphi$, there is a peak splitting at $\omega_q$ which prevails for small chemical potentials and only for $\varphi=0$, the Dirac cone approximation is recovered for $\mu\to0$.

\section{Acknowledgments}
We acknowledge useful discussions with G. G\'omez-Santos. This work has been supported by FCT under grant PTDC/FIS/101434/2008 and MIC under grant FIS2010-21883-C02-02.


\end{document}